 \documentclass[aps,floatfix,prd,twocolumn,showpacs,groupedaddress]{revtex4}
\usepackage{graphicx}

\newcommand{\bma}[1]{\mbox{\boldmath${#1}\/$}}
\newcommand{\Cdot}{\bma{\cdot}}
\newcommand{\Nabla}{\bma{\nabla}}

\begin{document}

\title[Theory and modeling of the magnetic field\ldots]{Theory and
modeling of the magnetic field measurement in LISA PathFinder}

\author{M. D\'\i az-Aguil\'o$^{1,2}$} 
\email[]{marc.diaz.aguilo@fa.upc.edu}
\author{E. Garc\'\i a-Berro$^{1,2}$}
\affiliation{$^1$Departament de F\'\i sica Aplicada, 
                 Universitat Polit\`ecnica de Catalunya, 
                 c/Esteve Terrades, 5, 
                 08860 Castelldefels, 
                 Spain\\
             $^2$Institut d'Estudis Espacials de Catalunya,
                 c/Gran Capit\`a 2--4, 
                 Edif. Nexus 104, 
                 08034 Barcelona, 
                 Spain}
\author{A. Lobo$^{3,2}$}
\affiliation{$^3$Institut de Ci\`encies de l'Espai, CSIC, 
                 Campus UAB, Facultat de Ci\`encies, Torre C-5, 
                 08193 Bellaterra, 
                 Spain}

\date{\today}

\begin{abstract}
The  magnetic  diagnostics  subsystem  of the  {\sl  LISA}  Technology
Package  ({\sl  LTP})  on   board  the  {\sl  LISA}  PathFinder  (LPF)
spacecraft includes  a set  of four tri-axial  fluxgate magnetometers,
intended to  measure with high  precision the magnetic field  at their
respective positions. However, their  readouts do not provide a direct
measurement of the magnetic field at the positions of the test masses,
and hence an interpolation method  must be designed and implemented to
obtain the values of the  magnetic field at these positions.  However,
such  interpolation process faces  serious difficulties.   Indeed, the
size  of   the  interpolation  region   is  excessive  for   a  linear
interpolation to be  reliable while, on the other  hand, the number of
magnetometer channels  does not provide  sufficient data to  go beyond
the  linear  approximation.   We  describe an  alternative  method  to
address this  issue, by means  of neural network algorithms.   The key
point in this approach is the ability of neural networks to learn from
suitable  training  data representing  the  behavior  of the  magnetic
field.  Despite the relatively  large distance between the test masses
and the  magnetometers, and the insufficient number  of data channels,
we find that our artificial neural network algorithm is able to reduce
the estimation errors  of the field and gradient  down to levels below
10\%, a  quite satisfactory result.   Learning efficiency can  be best
improved  by making  use of  data obtained  in  on-ground measurements
prior to  mission launch  in all relevant  satellite locations  and in
real operation conditions.  Reliable information on that appears to be
essential for  a meaningful assessment  of magnetic noise  in the~{\sl
LTP}.
\end{abstract}


\maketitle


\section{Introduction}
\label{chap.1}

{\sl  LISA}  Pathfinder  ({\sl  LPF})  is  a  science  and  technology
demonstrator  programmed  by the  European  Space  Agency ({\sl  ESA})
within its  {\sl LISA} mission  activities~\cite{bib:LISA}. {\sl LISA}
(Laser Interferometer Space Antenna) is a joint {\sl ESA-NASA} mission
which will  be the first  low frequency (milli-Hz)  gravitational wave
detector,   and  also   the  first   space-borne   gravitational  wave
observatory. {\sl LPF}'s payload  is the {\sl LISA} Technology Package
(LTP), and will be the  highest sensitivity geodesic explorer flown to
date. The  LTP is designed  to measure relative  accelerations between
two test  masses (TM)  in nominal free  fall (geodesic motion)  with a
noise budget of
\begin{equation}
 S^{1/2}_{\delta a, LPF}(\omega) \leq 3 \times 10^{-14} \left[ 1 +
 \left(\frac{\omega/2\pi}{3\;
\rm{mHz}}\right)^2 \right]  \frac{\rm{m\; s}^{-2}}{\sqrt{\rm Hz}}
\label{eq.0}
\end{equation}
in the frequency band between 1 mHz and 30 mHz~\cite{bib:trento}.

Noise  in   the  {\sl  LTP}   arises  as  a  consequence   of  various
disturbances,  mainly generated  within the  spacecraft  itself, which
limit  the   performance  of  the  instrument.   A   number  of  these
disturbances  are  monitored  and  dealt  with by  means  of  suitable
devices,    which     form    the    so-called     {\sl    Diagnostics
Subsystem}~\cite{bib:ere2006}. In {\sl LPF}, this includes thermal and
magnetic  diagnostics,  plus  the  radiation monitor,  which  provides
counting and  spectral information  on ionizing particles  hitting the
spacecraft.  The  magnetic diagnostics system  will be the  subject of
our attention here.

\begin{figure}
 \centering
 \includegraphics[width=0.9\columnwidth]{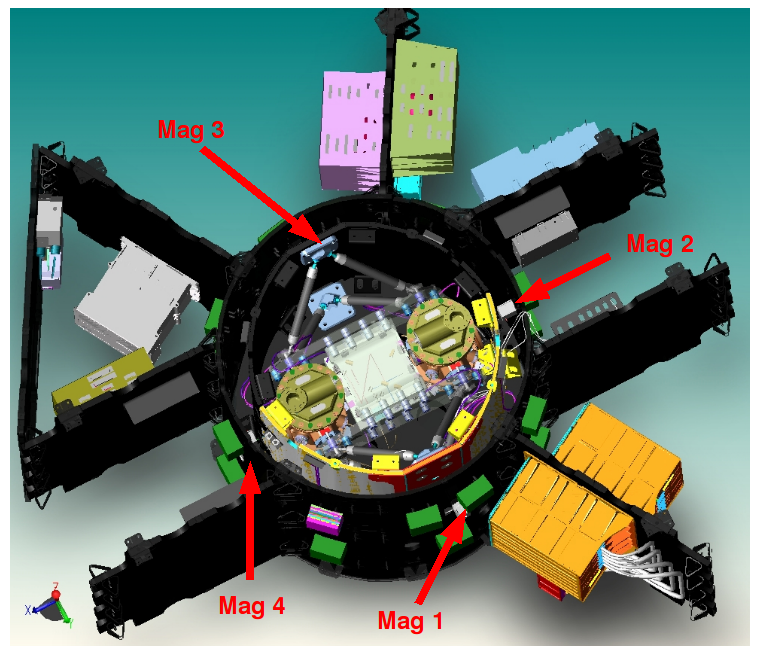}
 \caption{Artist view of the {\sl LPF} science-craft. The {\sl LCA} is
          in   the  center,   surrounded  by   a   double  cylindrical
          shield.  Outside  it,  a  number  of  electronic  boxes  are
          represented,   most  of  which   are  sources   of  magnetic
          field.  The four  magnetometers are  the white  little boxes
          indicated  by the  arrows (magnetometer  \#3 is  however not
          visible), and are mounted on the outer cylindrical shell.
 \label{fig.1}}
\end{figure}

One of the most important functions of the LTP magnetic diagnostics is
the  determination of  the  magnetic  field and  its  gradient at  the
positions of the  TMs.  For this, it includes a  set of four tri-axial
fluxgate magnetometers,  intended to  measure with high  precision the
magnetic field at the positions  they occupy in the spacecraft --- see
figure~\ref{fig.1}.  Their  readouts do  not however provide  a direct
measurement of the magnetic field  at the positions where the TMs are,
and an interpolation method must therefore be implemented to calculate
it. In the circumstances we  face, this is a difficult problem, mostly
because the  magnetometers layout  is such that  they are  too distant
from the locations of the TMs  compared with the typical scales of the
distribution  of magnetic sources  in the  satellite. Its  solution is
however    imperative    since    magnetic    noise    can    be    as
high  as  40\,\%  of   the  total  budget~\cite{bib:trento}  given  by
Eq.~(\ref{eq.0}), and hence it must be properly quantified.

In order to design a suitable interpolation scheme, information on the
actual distribution  of magnetic sources  is necessary. Data  from the
spacecraft manufacturer (EADS Astrium  Stevenage, UK) have kindly been
handed to  us~\cite{bib:wealthy} for this purpose.  According to these
data, magnetic sources can be characterized as magnetic dipoles, whose
positions  are known  and whose  magnetic  moments are  only known  in
modulus --- not  in orientation. Most of these  dipoles are associated
to  electronic boxes,  with  a few  genuinely  magnetic elements.   An
exception to  this rule  is the solar  panels, which cover  the entire
spacecraft and  can hardly be  considered as a  dipole as seen  by the
magnetometers.   They  are however  designed  so  that their cells are
arranged to minimize magnetic effects  by having their rim wires wound
contiguous and in opposite senses.

Astrium data are  based on system design, so  validation with the real
spacecraft must  be done  by means of  experiment, which is  of course
included in  the planned  activities before launch.  Actually, though,
the structure of the magnetic source distribution and their properties
will not  be directly  visible either to  the magnetometers or  to the
interpolation  algorithms, which  will just  work with  magnetic field
values no matter  how they are generated. Nevertheless,  we think that
the  information available so  far, though  not final,  qualifies very
well as a  guide to the elaboration of a magnetic  model which will be
needed to define and verify the performance of the analysis algorithms
which  will  eventually  be  applied  to the  data  delivered  by  the
satellite in flight.

In this paper we  will make use of the dipole model  of the sources to
assess  the  performance  of  two  different  types  of  interpolation
methods:  multipole interpolation and  neural network  algorithms. The
first is the more  immediate one to try, but as we  will show below it
is not  as efficient as one  might expect {\sl a  priori}. To overcome
this   problem  we   propose   a  novel   method,   based  on   neural
networks. Based  on the results  obtained with the same  dipole source
model,  our   solution  looks  promising  since  the   errors  of  the
interpolated fields and gradients are significantly smaller than those
obtained  with the  multipole approach.   The paper  is  structured as
follows. In Sect. \ref{chap.2} we provide a general description of the
problem.  It  follows  Sect.    \ref{chap.3},  where  we  discuss  the
multipole interpolation, whereas in  Sect. \ref{chap.4} we explain our
neural network  approach.  The results of applying  this algorithm are
presented  in Sect.   \ref{chap.5},  while in  Sect.  \ref{chap.6}  we
summarize our major findings and we draw our conclusions.


\section{General description of the problem}
\label{chap.2}

Magnetic  noise  in the  {\sl  LTP} is  allowed  to  be a  significant
fraction     of    the     total    mission     acceleration    noise:
$1.2\times10^{-14}$\,m\,s$^{-2}$\,Hz$^{-1/2}$  can  be apportioned  to
magnetism,      i.e.,     40\,\%      of     the      total     noise,
3$\times$10$^{-14}$\,m\,s$^{-2}$\,Hz$^{-1/2}$,  see  Eq.~(\ref{eq.0}).
This   noise   occurs   because   the   residual   magnetization   and
susceptibility of  the TMs couple  to the surrounding  magnetic field,
giving rise to a force
\begin{equation}
 {\bf F} = \left\langle\left[\left({\bf M} +
 \frac{\chi}{\mu_0}\,{\bf B}\right)\Cdot\Nabla\right]{\bf B}\right\rangle V
 \label{eq.1}
\end{equation}
in each of  the TMs. In this expression {\bf B}  is the magnetic field
in  the TM,~$\chi$  and {\bf  M} are  its magnetic  susceptibility and
density of magnetic moment (magnetization), respectively, and $V\/$ is
the  volume  of the  TM;  $\mu_0$  is  the vacuum  magnetic  constant,
$4\pi\times 10^{-7}$\,m\,kg\,s$^{-2}$\,A$^{-2}$),  and $\langle \cdots
\rangle$  indicates  TM  volume  average  of  the  enclosed  quantity.
Moreover, the  magnetic field and  its gradient randomly  fluctuate in
the regions occupied by the  test masses, thus resulting in a randomly
fluctuating force:
\begin{equation}
 {\bf \delta F} = \left\langle\left[\left({\bf M} +
 \frac{\chi}{\mu_0}\,{\bf B}\right)\Cdot \delta \Nabla\right]{\bf B}+
 \frac{\chi}{\mu_0}\,\left[\delta{\bf B}\Cdot\Nabla\right]{\bf B}
 \right\rangle V
 \label{eq.2}
\end{equation}
where  $\delta{\bf  B}$ represents  the  fluctuation  of the  magnetic
field,  and   $\delta\Nabla$  stands   for  the  fluctuation   of  the
gradient~\cite{bib:ntcs}.

Quantitative  assessment of magnetic  noise in  the {\sl  LTP} clearly
requires  real-time monitoring of  the magnetic  field, which  in {\sl
LPF}  is  done   by  means  of  a  set   of  four  tri-axial  fluxgate
magnetometers~\cite{bib:DDS_LTP}.     These     devices     have     a
high-permeability magnetic  core, which drives a  design constraint to
keep them somewhat far from the TMs.  The price to be paid for this is
that the measured field is not  directly useful (we need to know it at
the positions of the TMs). Hence,  a procedure to estimate it at these
positions, based on  the data delivered by the  magnetometers, must be
set up.

\begin{figure}
\centering
\includegraphics[width=0.9\columnwidth]{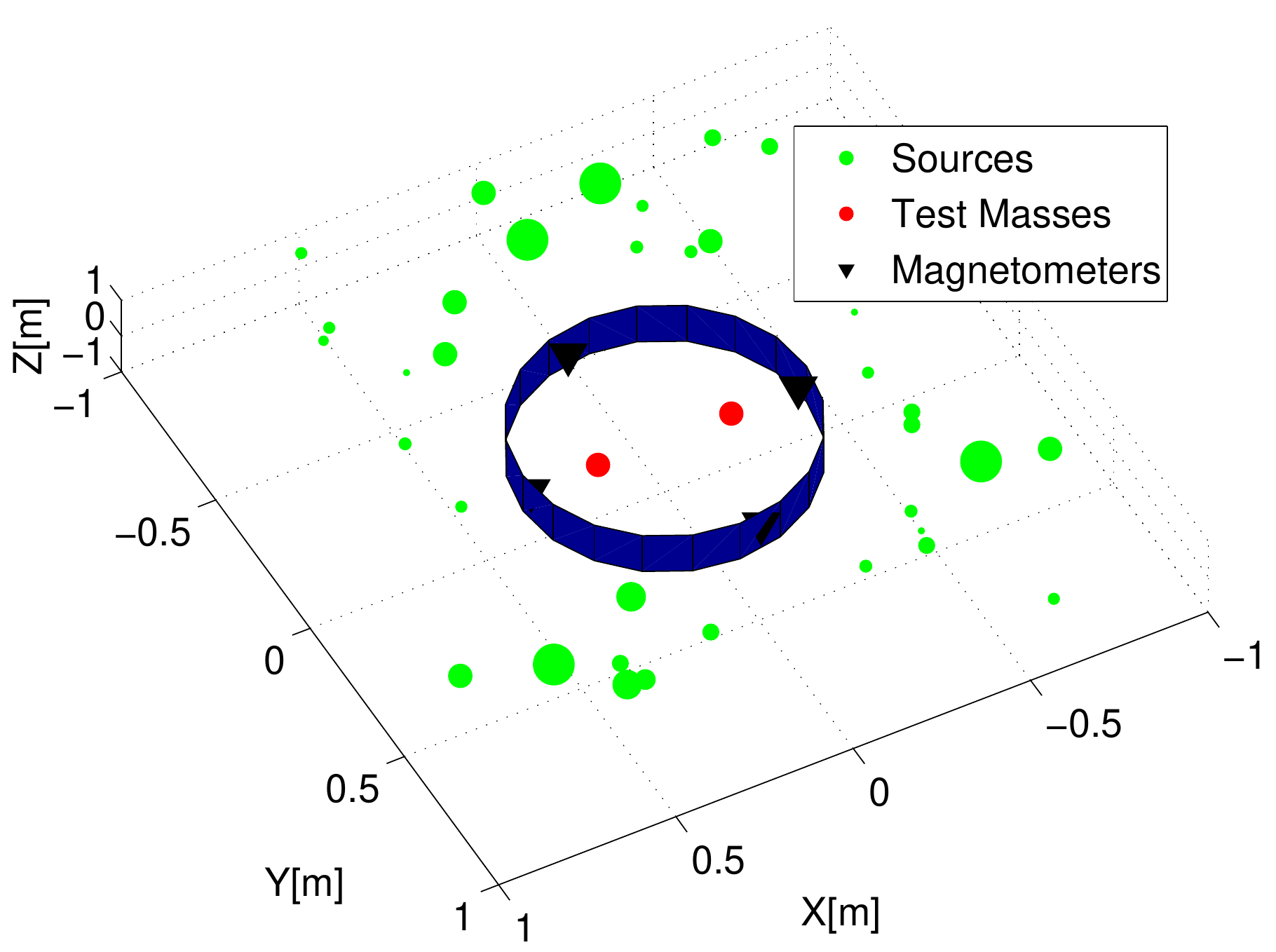}
\caption{Conceptual  diagram:  magnetic   sources  (green  dots,  size
         proportional  to the modulus  of the  magnetic moment  of the
         source), Test  Masses (red dots), and  the four magnetometers
         (black triangles).   Also represented  (in dark blue)  is the
         wall of the {\sl LCA}.
\label{fig.2}}
\end{figure}

As previously mentioned, the sources of magnetic field are essentially
electronics boxes plus a  few genuinely magnetic components inside the
spacecraft. The  interplanetary magnetic field is  orders of magnitude
weaker, hence of little relevance  to the effects considered here, and
solar    panel   effects    will   not    be   considered    ---   see
section~\ref{chap.1}. There  are no  sources of magnetic  field inside
the {\sl LTP} Core Assembly  ({\sl LCA}), all being placed outside its
walls.  The number of Astrium identified sources is around 40, and can
be    modeled    as    point   magnetic    dipoles~\cite{bib:wealthy}.
Figure~\ref{fig.2} gives an overview  of the geometry, see caption for
details.

\section{Multipole interpolation theory}
\label{chap.3}

Perhaps the most immediate  (and obvious) procedure to interpolate the
magnetic field is to resort  to its multipole structure. This is known
to     be     the     best     option     in     some     mathematical
sense~\cite{bib:jackson}. Consequently, we  first describe the details
of its implementation, and then we assess its practical merit.

We will treat the {\sl LCA}  region as a vacuum.  This is a reasonable
hypothesis,   as    the   materials   inside    it   are   essentially
non-magnetic. Accordingly, the magnetic  field has zero divergence and
rotational  \footnote{Given the  distances in  the spacecraft,  in the
order of 1\,m, propagation  effects will be neglected. Time dependence
will therefore be purely parametric, i.e., the time variable will just
label the value the field takes on at that time.}:
\begin{equation}
 \Nabla\Cdot  {\bf B}({\bf x},t) = 0\quad\quad
 \Nabla\times {\bf B}({\bf x},t) = 0
 \label{eq.4}
\end{equation}
Since $\Nabla\times {\bf B}({\bf x},t) = 0$, we thus have
\begin{equation}
  {\bf B}({\bf x},t) = \Nabla\Psi({\bf x},t)
 \label{eq.5}
\end{equation}
where  $\Psi({\bf x},t)$  is a  scalar function.   Additionally, since
$\Nabla\Cdot  {\bf B}({\bf x},t)$\,=\,0,  too, it  immediately follows
that $\Psi({\bf x},t)$ is a harmonic function, or
\begin{equation}
 \nabla^2\Psi({\bf x},t) = 0
 \label{eq.6}
\end{equation}

The solution to this equation can be expressed as an orthogonal series
of the form
\begin{equation}
 \Psi({\bf x},t) = \sum_{l=0}^\infty\sum_{m=-l}^l\,
 M_{lm}(t)\,r^l\,Y_{lm}({\bf n})
 \label{eq.7}
\end{equation}
where
\begin{equation}
 r\equiv |{\bf x}|\ ,\qquad {\bf n}\equiv {\bf x}/r
 \label{eq.8}
\end{equation}
are the spherical coordinates of the field point {\bf x}, whose origin
is by  (arbitrary) convention assumed  in the geometric center  of the
{\sl   LCA}.    Equation~(\ref{eq.7})   could   also   contain   terms
proportional to  $r^{-l-1}$, but these  have been dropped  because the
field cannot  diverge at the center  of the {\sl  LCA}.  Actually, the
expansion of Eq.~(\ref{eq.7}) is only valid in a region interior to the
closest  field source.  Finally,  the coefficients  $M_{lm}(t)$, which
will be  called multipole  coefficients in the  sequel, depend  on the
sources of magnetic field.

To   obtain  the   field  components   we  take   the   derivative  of
Eq.~(\ref{eq.7}) following Eq.~(\ref{eq.5}):
\begin{equation}
 {\bf B}({\bf x},t) = \Nabla\Psi({\bf x},t) =
 \sum_{l=1}^\infty\sum_{m=-l}^l\,
 M_{lm}(t)\,\Nabla[r^l\,Y_{lm}({\bf n})]
 \label{eq.9}
\end{equation}

According to standard mathematics, the coefficients $M_{lm}(t)$ can be
fully determined if the magnetic field is known at the boundary of the
volume where the field equations are considered, in this case the {\sl
LCA}.  This data is of course  not available to us, since we only know
{\bf  B} in  four  points  of the  boundary,  where the  magnetometers
are. Therefore the  question we need to address is:  how many terms of
the  series can  we possibly  determine on  the basis  of  the limited
information   available?   Or,   equivalently,   how  many   multipole
coefficients can  we estimate,  given the magnetometers  readout data?
Then,  also, to  which accuracy  can we  estimate the  actual magnetic
field  after the maximum  number of  multipole coefficients  have been
calculated?

The answer to the first  question above is actually not difficult: let
us assume  that the  series in Eq.~(\ref{eq.9})  is truncated  after a
maximum  multipole  index value  $l\/$\,=\,$L$.  The estimated  field,
${\bf B}_{\rm e}$, is then given by:
\begin{equation}
 {\bf   B}_{\rm e}({\bf   x},t)  =   \sum_{l=1}^L\sum_{m=-l}^l\,
 M_{lm}(t)\,\Nabla[r^l\,Y_{lm}({\bf n})]
 \label{eq.10}
\end{equation}
The number of terms in this sum is
\begin{equation}
 N(L) = \sum_{l=1}^L\,(2l+1) = L(L+2)
 \label{eq.11}
\end{equation}
which obviously equals the  number of multipole coefficients needed to
evaluate   the  sum.    For  example,   we  have   $N(2)\/$\,=\,8  and
$N(3)\/$\,=\,15.  On  the other hand, the number  of magnetometer data
channels is 12 --- three channels per magnetometer, as the devices are
tri-axial. This means we cannot  push the series beyond the quadrupole
($l\/$\,=\,2)  terms.  This  means that  since  we only  have 12  data
channels  we  have  some  redundancy  to  determine  the  first  eight
$M_{lm}(t)$  coefficients  up  to  $l\/$\,=\,2, though  we  also  lack
information  to evaluate  the  next seven  octupole terms  \footnote{A
clarification   is  in   order  here.    The   multipole  coefficients
$M_{lm}(t)$ are  actually complex numbers, which may  mislead one into
inferring that actually fewer can  be calculated.  This is however not
so  because  of  the  symmetry  $M_{lm}(t)$\,=\,$(-1)^mM^*_{l,-m}(t)$,
which ensures that ${\bf B}({\bf x},t)$ is actually a real number.}.

In order  to make a best  estimate of the  $M_{lm}(t)$, a least-square
method is set up as follows. Firstly, we define a quadratic error:
\begin{equation}
 \varepsilon^2(M_{lm}) = \sum_{s=1}^4\,\left|
 {\bf B}_{\rm r}({\bf x}_s,t) -
 {\bf B}_{\rm e}({\bf x}_s,t)\right|^2
 \label{eq.12}
\end{equation}
where ${\bf B}_{\rm r}$ is the real magnetic field and the sum extends
over the  number of magnetometers,  situated at positions  ${\bf x}_s$
($s\/$\,=\,1,\ldots,4).  We  then find those values  of $M_{lm}$ which
minimize the error:
\begin{equation}
 \frac{\partial\varepsilon^2}{\partial M_{lm}} = 0
 \label{eq.13}
\end{equation}

Once this  system of equations  is solved, the  estimated coefficients
$M_{lm}(t)$  are replaced  back  into Eq.~(\ref{eq.10})  and then  the
spatial arguments  {\bf x} substituted  by the positions of  each test
mass  to finally obtain  the interpolated  field values.  This process
needs  to  be  repeated  for  each  instant $t\/$  of  time  at  which
measurements  are taken,  thereby generating  the magnetic  field time
series.  The  gradient  is  estimated  by taking  the  derivatives  of
Eq.~(\ref{eq.10}):
\begin{equation}
 \left.\frac{\partial B_i}{\partial x_j}\right|_{\rm e}({\bf x},t) =
 \sum_{l=0}^L\sum_{m=-l}^l\,M_{lm}(t)\;
 \frac{\partial^2}{\partial x_i\partial x_j} 
 \left[r^l\,Y_{lm}({\bf n})\right]
 \label{eq.14}
\end{equation}

It is  to be  noted that Eq.~(\ref{eq.10})  is a polynomial  of degree
$L\/$$-$1 in the space  coordinates $(x,y,z)$, hence its degree equals
1  when  $L$\,=\,2.   Since  this  is  the most  we  can  get  of  the
magnetometer  readout channels,  the multipole  expansion  is actually
equivalent to  a {\sl linear}  interpolation of the field  between its
values at  the boundary  of the  {\sl LCA} and  its interior.   We may
therefore not expect this  method to produce excellent results, simply
because the magnetic field inside the  {\sl LCA} is weaker than at its
boundaries,  the reason  being  that the  magnetic  field sources  are
outside the  {\sl LCA}.  This  valley structure of the  magnetic field
needs at least octupole (quadratic) terms to be approximated, but this
would  require at  least one  more vector  magnetometer, which  is not
available.   By the  same argument,  the  field gradient  can only  be
approximated by  a {\sl constant}  value throughout the {\sl  LCA} ---
see Eq.~(\ref{eq.14}).

\subsection{Numerical simulations}
\label{chap.3.1}

In order to have a quantitative  idea of the actual performance of the
above interpolation  scheme, we make  use of the source  dipole model.
It has the following ingredients and assumptions:

\begin{enumerate}
 \item The  sources  of magnetic  are point  dipoles outside  the {\sl
       LCA}.
 \item The  sources  are  those identified  by  Astrium Stevenage,  as
       already mentioned,  whose positions in  the S/C are  known. The
       set itself, as  well as the source magnetic  parameters need to
       be  updated, but  the data  used (which  date back  to November
       2006) qualifies to verify  the performance of the interpolation
       methods.
 \item The  magnetic  field created  by the  dipole distribution  at a
       generic point {\bf x} and time $t\/$ is therefore given by
       \begin{equation}
        {\bf B}({\bf x},t) = \frac{\mu_0}{4\pi}\;\sum_{a=1}^{n}
        \frac{3\left[{\bf m}_a(t)\Cdot {\bf n}_a\right]\,{\bf n}_a - 
        {\bf m}_a(t)}{\left|{\bf x}-{\bf x}_a\right|^3}
       \label{eq.15}
       \end{equation}
       where      {\bf       n}$_a$\,=\,      $\left({\bf      x}-{\bf
       x}_a\right)$/$\left|{\bf x}-{\bf  x}_a\right|$ are unit vectors
       connecting  the the $a$-th  dipole {\bf  m}$_a$ with  the field
       point {\bf x}, and $n\/$ is the number of dipoles.
 \item Fluctuations of the dipoles, both in modulus and direction, are
       unknown,  but this  is not  essential to  assess  the numerical
       performance of the algorithm.
\end{enumerate}

We aim to compare interpolated  magnetic field results with exact ones
within  the context  and scope  of  the above  model. To  artificially
simulate  several possible scenarios,  we will  take advantage  of the
uncertainties in  the source dipole orientations  to randomly generate
different  magnetic field  patterns,  which we  intend to  reconstruct
based on the multipole expansion.  More specifically, the procedure is
the following one:

\begin{enumerate}
\label{model}
 \item Each dipole has a known fixed position in the spacecraft, and a
       fixed modulus,  also known. The  number of magnetic  dipoles is
       also fixed to 37, which is the number in Astrium's list.
 \item The orientations of the dipoles are instead unknown. An example
       scenario  is characterized by  a specific  selection of  the 37
       dipole orientations.
 \item In order  to explore the behavior of the  algorithm, a batch of
       examples  are  examined,   each  corresponding  to  a  randomly
       generated set of dipole orientations.
 \item In each case, Eq.~(\ref{eq.13}) is solved for $M_{lm}$, and the
       field   estimate   at  each   TM   is   then  calculated   with
       Eq.~(\ref{eq.10}).  In  the last  step, the result  is compared
       with the  theoretical one  given in Eq.~(\ref{eq.15}),  and the
       differences annotated.
 \item Finally, a statistical  analysis of the differences (errors) is
       done.
\end{enumerate}

The random character of the  procedure may seem unrealistic, since the
actual  satellite  configuration  is  not  random.  In  this  context,
however,  randomness  is  an   efficient  way  of  mimicking  lack  of
knowledge.  As we  will see  in the  next section,  numerical analysis
based  on this  methodology  sheds much  light  on the  merits of  the
interpolation procedure ---  as it will also be the  case when we come
to neural networks performance in section~\ref{chap.5}.

\begin{figure}[b]
\centering
 \includegraphics[width=0.4\columnwidth]{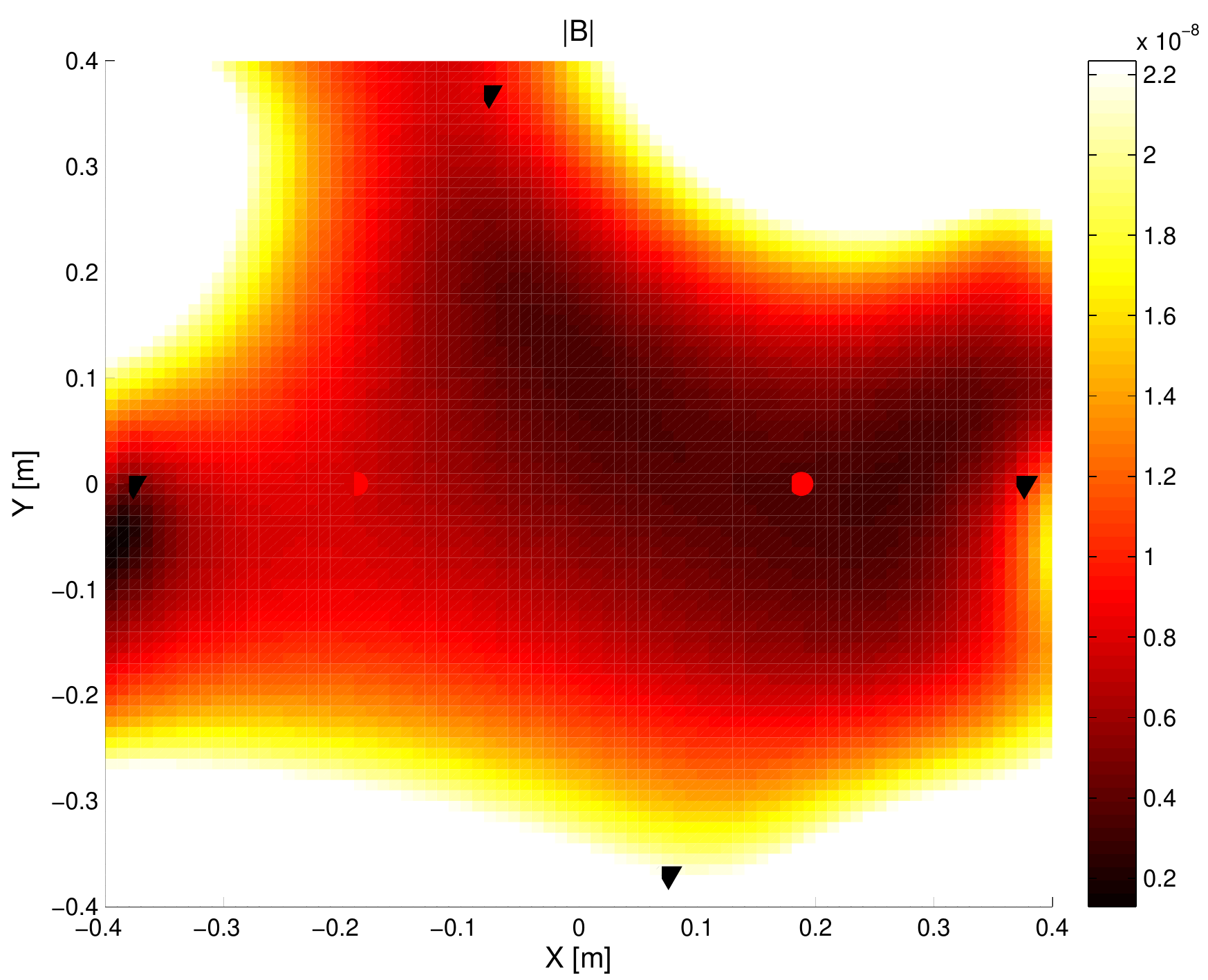}\quad
 \includegraphics[width=0.4\columnwidth]{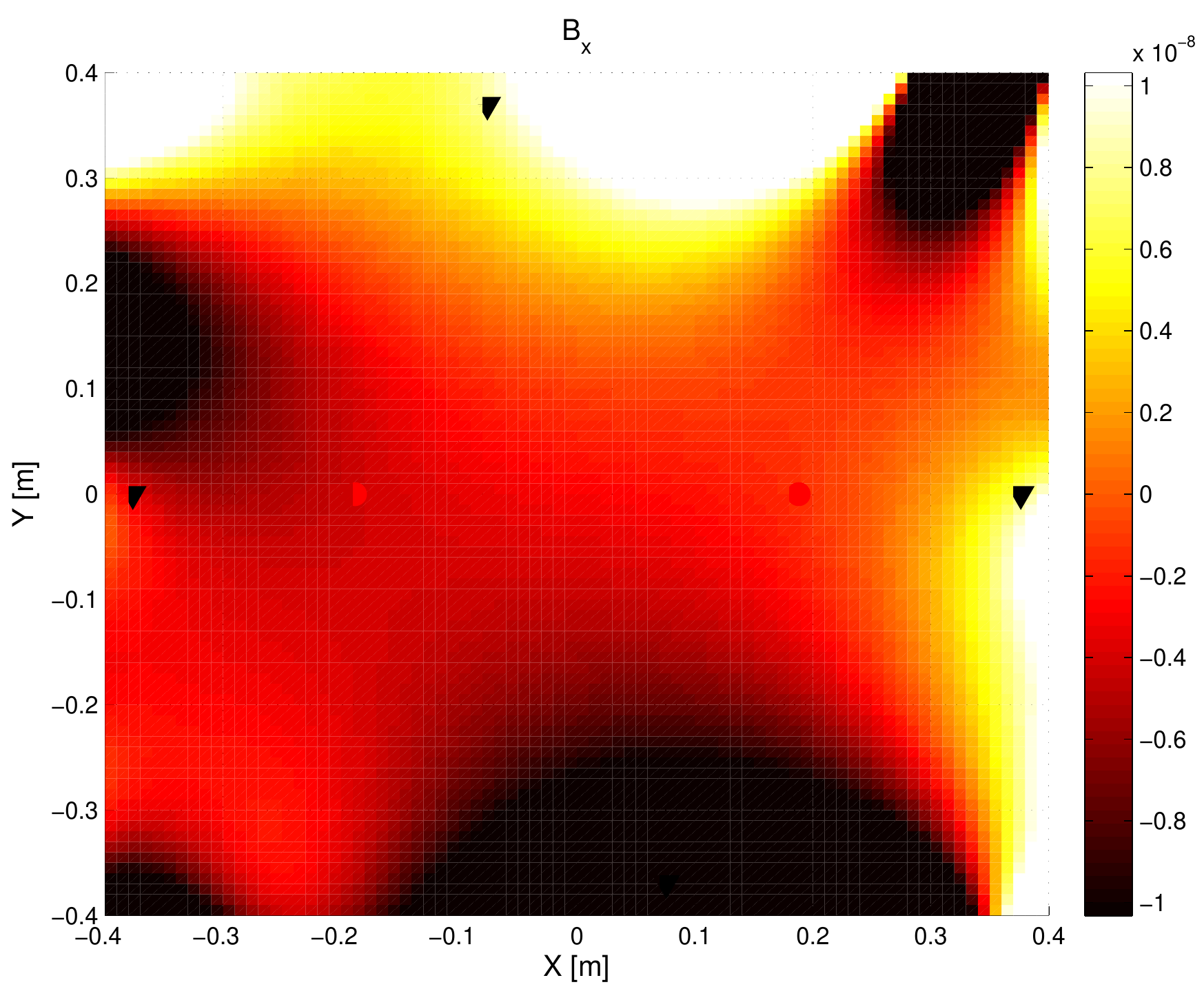} \\
 \includegraphics[width=0.4\columnwidth]{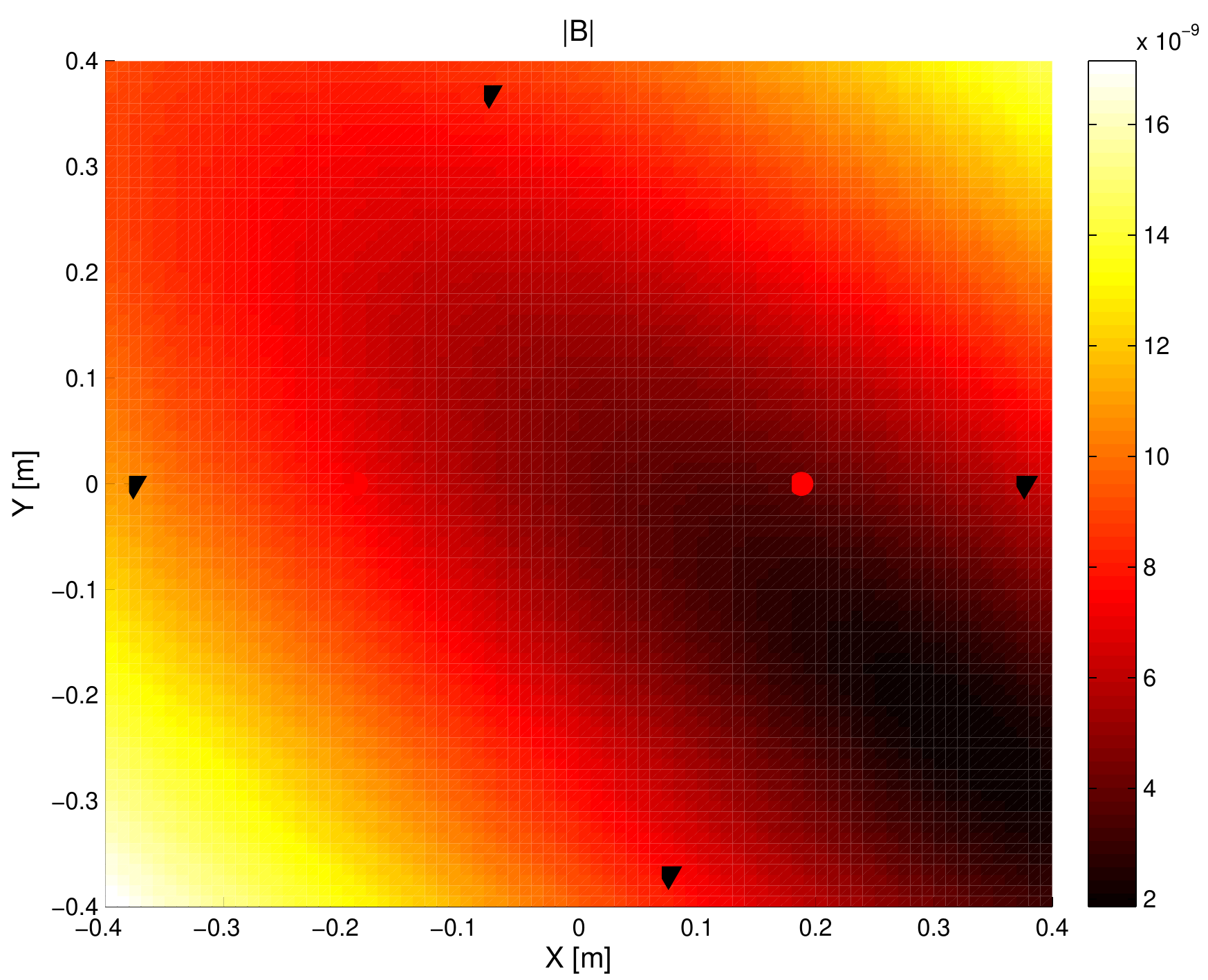}\quad
 \includegraphics[width=0.4\columnwidth]{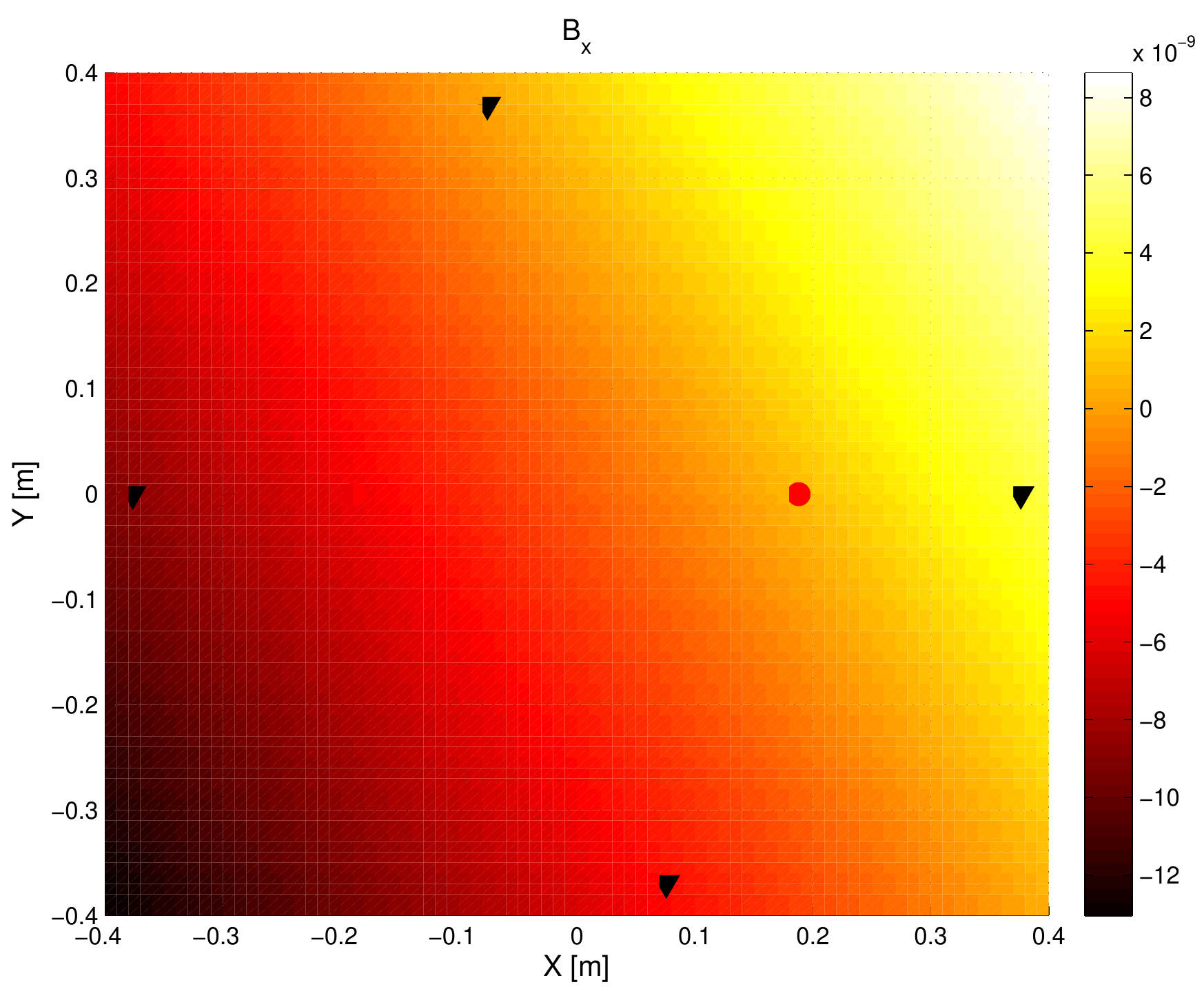} \\
 \caption{Magnetic field contour  plots in the {\sl LCA}  region for a
          given  source dipole configuration.  $x\/$ is  the direction
          between  the two  TMs, and  $y\/$ is  in  the ``horizontal''
          plane,  which in  the plot  is at  the TMs  centres  of mass
          altitude. Left  panels: moduli  of the magenetic  field. The
          top panel displays  the exact one, and the  bottom one shows
          its multipole  estimate.  Right panels: same as  in the left
          panels,  but for one  of the  field components  ($B_x$). The
          modulus of  the magnetic field shows a  complex structure in
          the central area, while $B_x\/$ has a saddle structure there
          --- see along the diagonals of the frame graph. The red dots
          mark the  centers of  the TMs, and  the black  triangles the
          positions of the magnetometers.
 \label{fig.2.1}}
\end{figure}

\subsubsection{Simulation results}
\label{chap.3.1.1}

In this section we summarise the most relevant results of the analysis
of   the  multipole   interpolation  method.   We  use   a   batch  of
1\,000~example  scenarios  such as  described  above. Magnetic  moment
orientations  were  chosen  by  randomly  picking values  of  the  two
defining  spherical  angles  $(\theta,\varphi)$ from  two  independent
uniform distributions.

Figure \ref{fig.2.1}  graphically represents  a magnetic field  map in
the {\sl  LCA} region corresponding  to an arbitrarily  chosen example
out of  the 1\,000 considered. The  valley structure is  very clear in
the $|{\bf B}|$ plot, while the $B_x\/$ component shows a saddle shape
--- see figure caption. $B_y\/$ and $B_z\/$ show qualitatively similar
forms, and  thus we  do not  show them.  The  elliptical forms  in the
estimate of  $|{\bf B}|$ are due  to the quadratic  combination of the
field  components.  The  estimate of  $B_x\/$ shows  instead  a linear
structure,  with  constant  gradient  in  all  directions.  Naked  eye
inspection  immediately reveals a  poor resemblance  between estimated
and exact quantities, but let us elaborate some numerical data.

\begin{figure}[t]
 \centering
 \includegraphics[width=0.4\columnwidth]{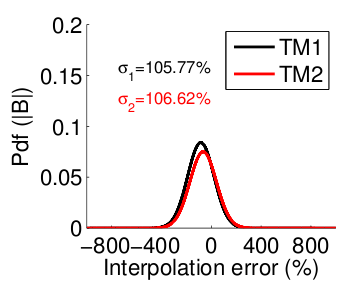}\quad
 \includegraphics[width=0.4\columnwidth]{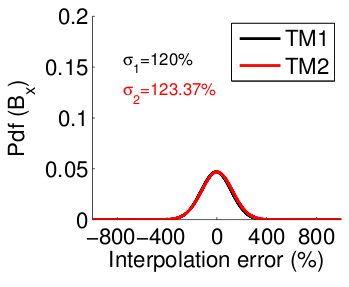}
 \caption{Binned errors of the estimated modulus of the magnetic field
          and   of   its  $x\/$ component.   They   are  reported   in
          percentage. Colors correspond to  each of the {\sl LTP} TMs,
          respectively.  Inset  values of  $\sigma$  indicate the  rms
          half-width  of the distributions.  Solid lines  are Gaussian
          fits to the histograms.
\label{fig.3.1}}
\end{figure}

Figure~\ref{fig.3.1}  displays the  binned distribution  of estimation
errors, defined by
\begin{equation}
 \varepsilon(|{\bf B}|) = \frac{|{\bf B}_{\rm e}|
 - |{\bf B}_{\rm r}|}{|{\bf B}_{\rm r}|}\ ,\quad
 \varepsilon(B_x) = \frac{B_{x,{\rm e}}
 - B_{x,{\rm r}}}{|{\bf B}_{\rm r}|}
 \label{eq.16}
\end{equation}
where   we  have   used   a  denominator   $|{\bf   B}_{\rm  r}|$   in
$\varepsilon(B_x)$  to avoid  meaningless infinities  when  $B_x\/$ is
close to  zero. $B_y\/$  and $B_z\/$ show  similar trends and  are not
displayed.   As can  be seen,  errors average  to zero,  but  have rms
deviations well  above 100\,\%. Even worse,  outliers are significant,
as  can be  seen  in Table~\ref{tab:table1},  where averaged  absolute
values  over  the  1\,000   simulated  cases  are  displayed.  Except,
obviously, for the modulus error,  we are around 500\,\%, but detailed
examination of  individual data further  shows that errors as  high as
2\,000\,\% eventually happen.

\begin{table}
\caption{\label{tab:table1}Averaged  absolute value of  the estimation
         errors in  the components  of the magnetic  field and  of its
         modulus. They are reported in relative percent.}
\begin{ruledtabular}
\begin{tabular}{lcc}
  & TM1 & TM2 \\
\hline
$\varepsilon(B_x)$       & 493.7 & 640.4 \\
$\varepsilon(B_y)$       & 330.5 & 543.1 \\
$\varepsilon(B_z)$       & 359.5 & 368.2 \\
$\varepsilon(|\bf{B}|)$  &  88.6 &  75.7 \\
\end{tabular}
\end{ruledtabular}
\end{table}

The most  salient features  of the numerical  analysis can  be briefly
summarized.  Firstly we find that magnetic field estimation errors are
very variable, ranging  from very few percent to  over 1\,000\,\% and,
secondly,  these huge uncertainties  happen in  an utterly  random and
fully unpredictable  way.  The {\sl a posteriori}  conclusion is quite
simple: the intrinsic linear  character of the interpolation scheme is
not capable  of reproducing the  field structure inside the  {\sl LCA}
--- hence at the positions of  the TMs --- and, therefore, can produce
very good  or very bad  results just by  accident.  In addition to not
being  predictable, the  average error  is  any case  too large.   The
ultimate reason for  such poor performance is the  the small number of
magnetometers as  well as  their positioning: four  magnetometers only
allow for  a field multipole  expansion up to quadrupole  terms, which
means that the field values  at the TMs are just linearly interpolated
between magnetometer readouts at the boundary of the {\sl LCA}. On the
other hand, the magnetometers are closer to the magnetic field sources
than they  are to  the TMs, which  prevents resolution of  the spatial
field structure details inside the {\sl LCA} with only linear terms in
the space coordinates.


\begin{figure*}[t]
 \centering
 \includegraphics[width=0.7\textwidth]{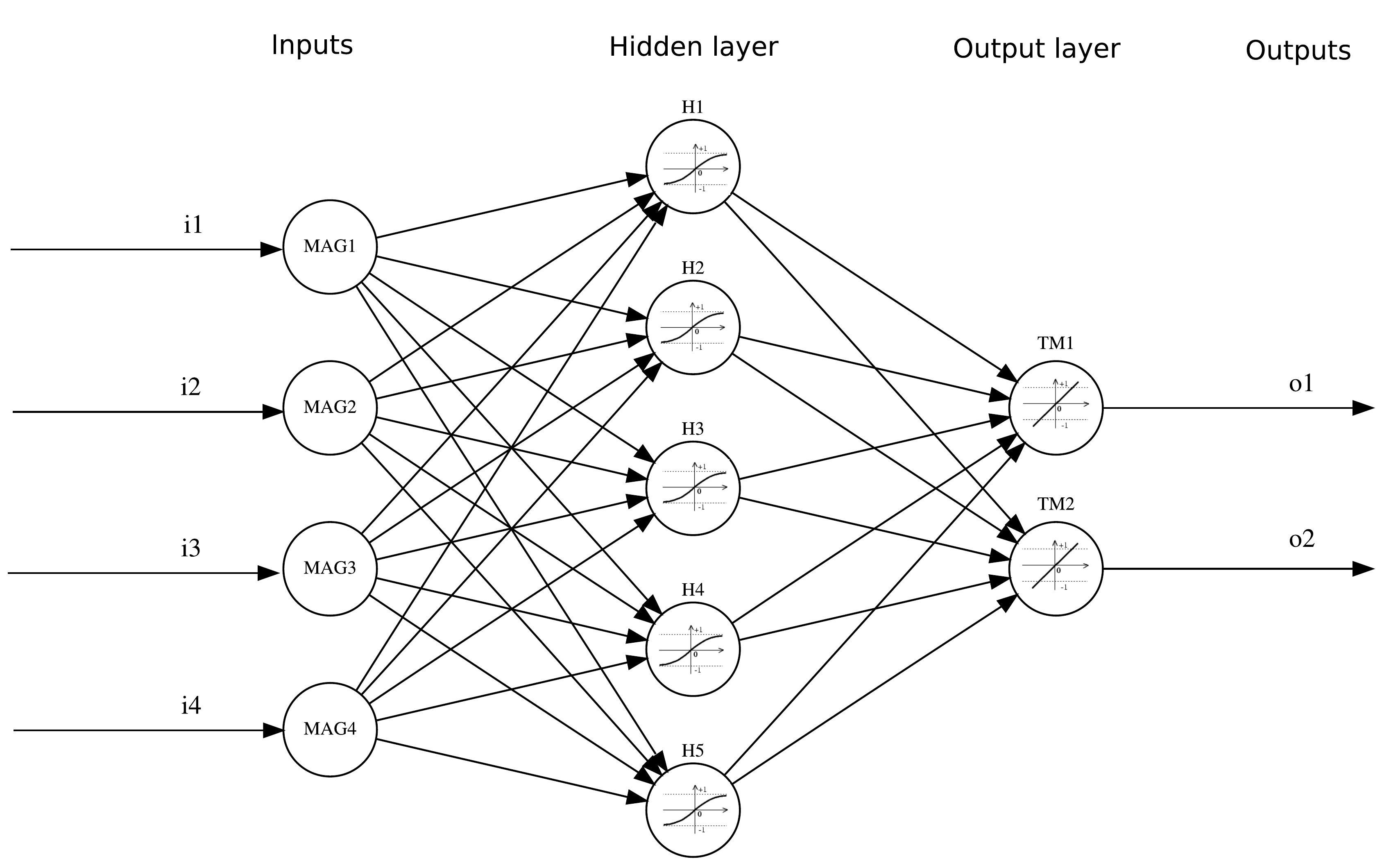}
 \caption{Feed-forward  neural   network  architecture.  Magnetometers
          readings are  the system inputs, and estimates  of the field
          and gradient  at the  positions of the  test masses  are the
          outputs  of  the  system.  In this  architecture,  only  one
          intermediate,  or  hidden layer  is  assumed.   Each of  the
          circles represents  one neuron and corresponds  to the model
          of Eqs.~(\ref{eq.18}) and~(\ref{eq.18a}).}
 \label{fig.4}
\end{figure*}

\section{A novel approach: neural networks}
\label{chap.4}

Search for an alternative  approach to the above interpolation schemes
is imperative, otherwise the information provided by the magnetometers
will hardly  be useful  for the  main goal of  the {\sl  LTP} magnetic
diagnostics system, i.e., to quantify the contribution of the magnetic
noise  to the  total system  noise.  Here some  promising results  are
presented on the implementation of a completely different methodology:
neural networks~\cite{bib:Kecman}.

Artificial neural  networks are made up  of interconnecting artificial
neurons   (programming  constructs  that   mimic  the   properties  of
biological neurons)  that have the  capacity to learn  from processing
data.    Neural  networks   are  often   used  in   solving  nonlinear
classification and  regression tasks by learning from  data, hence are
worth trying with the present problem \cite{bib:neuralPaper}.

There are four sets of tasks which need to be implemented when solving
a problem with artificial neural networks:
\begin{enumerate}
 \item Neuron model selection
 \item Model and architecture selection
 \item Learning paradigm and learning algorithm selection
 \item Performance assessment
\end{enumerate}
We next go through them, one by one.

\subsection{Neuron model}
\label{chap.4.1}

The neuron is the basic unit of any neural network. It performs
the following two operations:

\begin{itemize}
 \item It collects  the inputs from all other  neurons connected to it
       and computes  a weighted sum  of the signals the  latter inject
       into it, generally  adding a bias as well.  If we represent the
       inputs by  a vector {\bf  x}\,$\equiv$\,$(x_1,\ldots,x_n)$, and
       the  weights by  a  {\bf w}\,$\equiv$\,$(x_1,\ldots,w_n)$  then
       this operation consists in calculating the sum
       \begin{equation}
         \Sigma = w_0 + \sum_{k=1}^n\,w_k x_k \equiv w_0 + {\bf w}^T {\bf x}
         \label{eq.18}
       \end{equation}
       where the superindex $T\/$ stands for transpose matrix; in this
       case, {\bf  w}$^T$ is a  row vector while  {\bf x} is  a column
       vector, so  that ${\bf w}^T {\bf  x}$ is the  scalar product of
       {\bf w}  and {\bf x}.  A term  $w_0$ is added to  form the most
       general linear function  of the vector argument {\bf  x}; it is
       called the bias.
 \item The  above  sum  is  used  as  the  argument to  the  so-called
       activation  function, $\varphi(\Sigma)$.  The  neuron's output,
       also known as its activation, is thus
       \begin{equation}
        o = \varphi(\Sigma)
        \label{eq.18a}
       \end{equation}
       In general, $\varphi(\Sigma)$ can be selected in many different
       ways.  Here, differentiable  activation functions will be used,
       which suit well  the gradient descent back-propagation learning
       algorithm --- see sections below.
\end{itemize}

\subsection{Neural network architecture}
\label{chap.4.2}

Artificial neural networks are software or hardware models inspired by
the structure and behavior of biological systems, and they are created
by a  set of neurons distributed  in layers. There  are many different
types  of neural  networks in  use today,  but the  architecture  of a
so-called feed-forward network, where  each layer of neurons is linked
with the next by means of a set of weights, is the most commonly used,
and will also be used here.  The specific architecture adopted in this
study is shown in figure~\ref{fig.4}. The data streams coming from the
magnetometers  will be  considered the  system inputs,  while magnetic
field results and their gradients  at the positions of the test masses
will be the system outputs.

\subsection{Learning paradigms and learning/training algorithms}
\label{chap.4.3}

The investigation of learning  algorithms is currently an active field
of  research. The design  and implementation  of an  adequate training
scheme  is  the  essential  ingredient for  obtaining  a  good-quality
estimate of the magnetic field and its gradient at the {\sl LTP} TMs.

\subsubsection{Learning paradigms}
\label{chap.4.3.i}

There  are  two major  learning  paradigms,  each  corresponding to  a
particular  abstract   learning  task.   These   are  {\sl  supervised
learning} and {\sl unsupervised learning}.

\begin{enumerate}
 \item Supervised learning. The idea of this paradigm is quite clearly
       suggested by  its very name. A  set of examples  is filed, each
       set  consisting   in  a  number   of  vector  of   inputs  (the
       magnetometers'  readouts in  this case)  and  the corresponding
       values of the magnetic field and  its gradient at the TMs for a
       given distribution  of dipoles in the spacecraft.   Let {\bf x}
       represent a  generic input vector,  and {\bf y}  the associated
       vector output. These two vectors constitute an example. The set
       of  filed examples  for supervised  learning is  thus a  set of
       pairs  ({\bf  x},{\bf y}),  where  {\bf  x}$\in$$X\/$ and  {\bf
       y}$\in$$Y\/$,  $X\/$  and  $Y\/$  being  some  suitable  sample
       spaces.
       The network is  then fed the inputs {\bf x}  of one example and
       let it  work out an output,  {\bf o}, say. This  output is then
       compared  with  the correct  one,  {\bf  y},  and an  error  is
       calculated  if {\bf o}\,$\neq$\,{\bf  y}.  Iterations  are then
       triggered to  adjust the weighting factors such  that the error
       is minimized. These will however vary as different examples are
       run, so a cost function is defined which enables the network to
       optimise the  set of  weights which works  best for the  set of
       examples analyzed, based on some suitable criterion.
 \item Unsupervised learning. In unsupervised learning a cost function
       is  to be  minimized  as well,  but  this function  can be  any
       relationship between  {\bf x} and the network  output, {\bf o},
       but  never taking into  account the  real expected  target. The
       cost     function     is     determined     by     the     task
       formulation.   Unsupervised  learning   is  thus   a   form  of
       self-adaptive  system, whose  guide is  not an  {\sl  a priori}
       knowledge  of  the  final  result  but  knowledge  gained  from
       experience.
\end{enumerate}

In either case,  the learning process is based  on the architecture of
the   network,  i.e.,  number   of  neurons   and  layers   and  their
interconnections, as  well as on  the activation functions.  These are
parameters which,  at least in the  simplest cases, are  tuned {\sl ab
initio} by the user based  on observed performance of the network.  In
this  study, supervised  learning  has been  the implemented  learning
paradigm.

\subsubsection{Learning algorithms}
\label{chap.4.3.ii}

There are many algorithms  for training neural networks. When training
feed-forward   neural    networks   with   supervised    learning,   a
back-propagation algorithm  is usually  implemented. The error  of the
mapping at the output is propagated backwards in order to readjust the
weights  and improve  the output  error for  the next  iteration.  The
propagation can be implemented  with different methods, the {\sl Ideal
Gradient Descent} being  a classic which will also  be used here, with
slight modifications that make the algorithm convergence faster.

Iterations on  the weights of  the different neurons at  the different
layers proceed according to the following algorithm:
\begin{equation}
 {\bf w}_{n+1} = {\bf w}_{n} -
 \eta \left.\frac{\partial E}{\partial {\bf w}} \right|_n
 \label{eq.19}
\end{equation}
where $n\/$  labels the  current iteration step,  and $\eta\/$  is the
learning rate, adjustable  by the user. $E\/$ is the  sum over the set
of training examples of the square errors of the outputs:
\begin{equation}
 E = \sum_{s}\;({\bf o} - {\bf y})^T\,({\bf o} - {\bf y})
 \label{eq.140}
\end{equation}
where $s\/$ stands for the number of examples, {\bf  o} is the (vector)
output  from the  network, while  {\bf y}  is the  target,  or correct
output in  the corresponding example.  The quantity $E\/$ can  only be
defined in supervised  learning, of course, and the  idea of the above
procedure is  to find  that point  in weight space  where $E\/$  is an
absolute minimum. $E\/$ can  therefore be considered the cost function
to be  minimized in this  particular supervised training  scheme, also
known as  {\sl batch mode} as  the analysis is done  across the entire
set of training patterns in a single block.

There are a  number of technical issues in  pursuing the iterations in
Eq.~(\ref{eq.19}), such as  the choice of the initial  set of weights,
the   identification  of   local   minima  of   $E\/$,  the   boundary
effects,\ldots which  need to be  addressed in each specific  case. We
skip a detailed  discussion of these matters here and  we focus on the
results obtained using our method.  For further details, the reader is
referred to Refs.~\cite{bib:Kecman} and \cite{bib:neuralPrunning}.

\subsection{Performance assessment}
\label{chap.4.4}

In this  last step, the trained  network must be  tested with examples
which differ from  those used in the learning  process. This is needed
to  assess  whether or  not  the trained  neural  network  is able  to
generate the expected results  when fed with previously unseen inputs,
hence determine its usability for the specific purpose it is intended.


\begin{figure}[t]
 \centering
 \includegraphics[width=0.8\columnwidth]{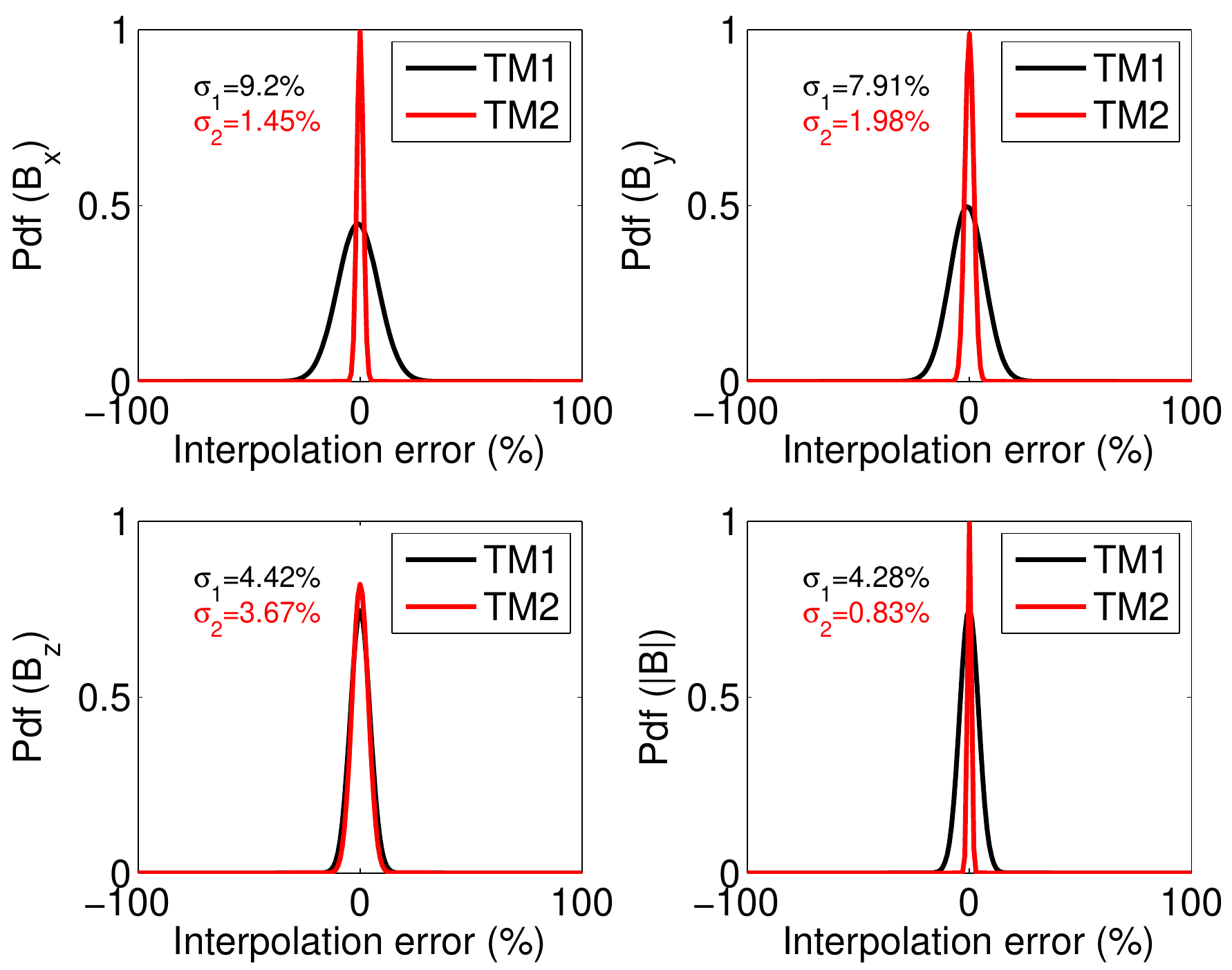}
 \caption{Error distributions for each field component at the position
          of test mass~1 (black line)  and test mass~2 (red line). The
          top left panel displays the results for $B_x$, the top right
          panel  shows the  error  distribution for  $B_y$, while  the
          bottom  left  panel depicts  the  distribution obtained  for
          $B_z$ and the bottom right panel that for $|{\bf B}|$.
 \label{fig.5}}
\end{figure}

\section{Results}
\label{chap.5}

Training  and  testing  have   been  done  based  on  different  field
realizations,  using the  same  model of  sources  and magnetic  field
described  in section~\ref{chap.3.1},  i.e., each  {\sl  example} will
consist in  the magnetic field  at the magnetometers'  positions, plus
the  magnetic field  and gradient  at the  TM positions,  all  of them
corresponding to a given configuration of the 37 Astrium dipoles.

Two different batches of  examples, each including 1\,000 realizations
of a possible magnetic  environment, have been generated following the
directives explained  in section~\ref{chap.3.1}.  The  first batch has
been used as  the training set for a neural network  with 12 inputs (3
inputs  for  each  of  the  4 vector  magnetometers)  and  16  outputs
representing the  field information  at the position  of the  two test
masses   (3  field   plus  5   gradient  components   per   test  mass
\footnote{Note  that only  5 of  the 9  gradient  components $\partial
B_i/\partial x_j$ are independent.   This is because the conditions of
Eq.~(\ref{eq.4}) imply that $\partial B_i/\partial x_j$ is a traceless
and  symmetric  matrix.}).   The   second  batch  has  been  used  for
validation to assess the performance of the network in front of unseen
magnetometers readings.

\subsection{Field estimation}
\label{chap.5.1}

Figure~\ref{fig.5}  shows  the  distribution  of relative  errors  (in
percentage) of the  estimated components of the magnetic  field at the
positions of each  TM. The plot is based on the  results of the 1\,000
validation runs described in the previous section. As can be observed,
the order of  magnitude of the errors of the  estimated fields are now
within  much  more  acceptable  margins  (below  $\sim  15\%$).   This
represents a reduction of estimation  errors of more than one order of
magnitude in comparison with the multipole expansion method.

During the training process, the neural network eventually learns that
the  magnetic  field  at  the   TMs  is  generally  smaller  than  the
magnetometers read --- with occasional  exceptions due to the rich and
complex  structure  of the  field  inside  the  {\sl LCA},  see  e.g.\
figure~\ref{fig.2.1}.   The  neural  network  is  able  to  derive  an
inference procedure which is actually  quite efficient, and it does so
by  proper adjustment  of its  weight matrix  coefficients {\bf  w} as
explained in section~\ref{chap.4.3.i}.   In order to better understand
the reaction of the trained neural network to the magnetometers' data,
we found instructive and  expedient to look into relationships between
the data  read by the  magnetometers and the magnetic  field estimates
generated at the output of  the neural network.  We chose to calculate
correlation  coefficients  between  input  and output  data,  and  the
results  are displayed in  Figure~\ref{fig.5.1}.  The  following major
features are identified:

\begin{figure}[t]
 \centering
 \includegraphics[width=0.8\columnwidth]{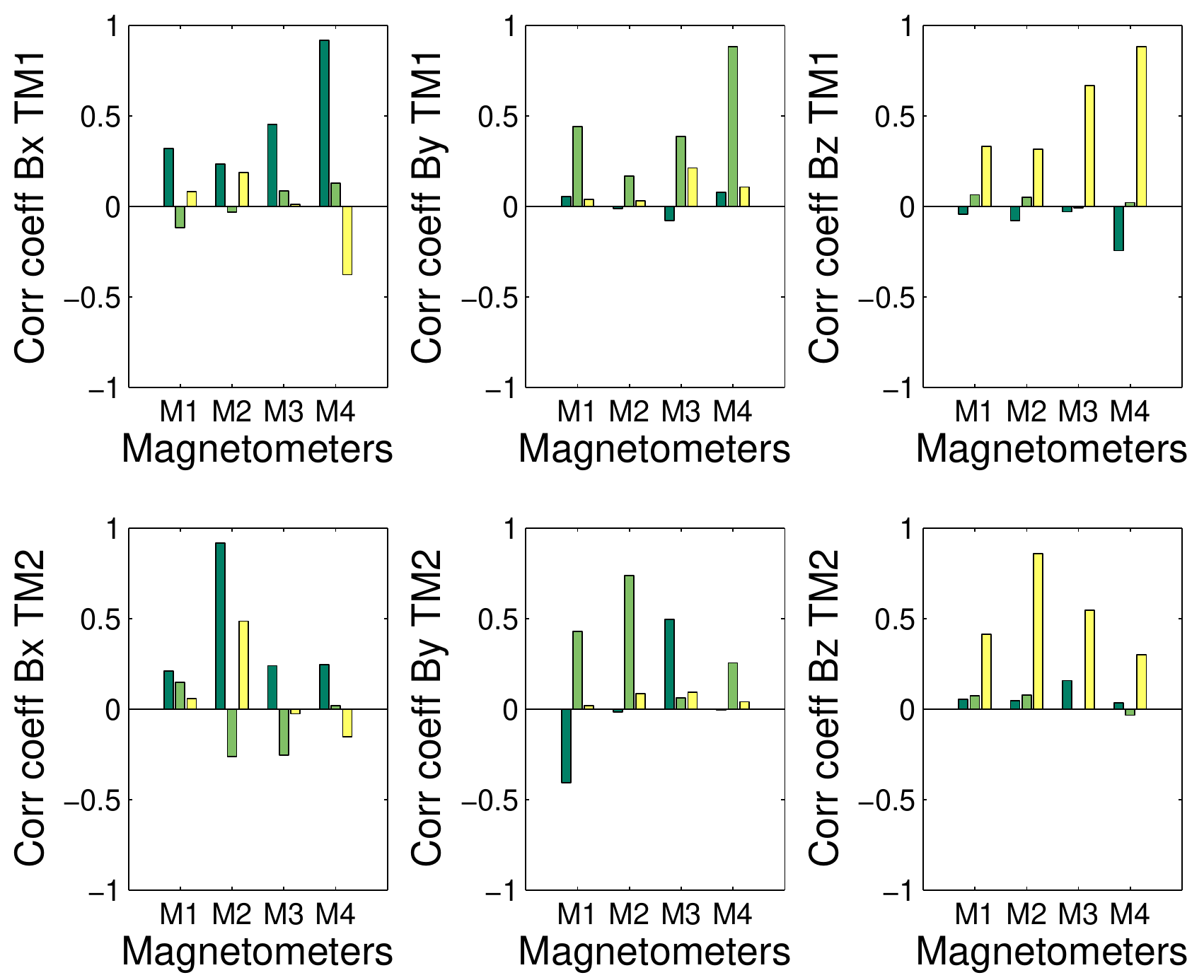}
 \caption{Correlation coefficient between the information entering the
          network (each  magnetometer input) and  the outputs provided
          by the trained network for the field estimates.
 \label{fig.5.1}}
\end{figure}

\begin{itemize}
 \item Each  component of  the field is  basically estimated  from the
       magnetometers reading  of the same component.  For example, the
       interpolation of the $B_x\/$ component in test mass 1 is mostly
       dependent on the $B_x\/$ readings of the magnetometers.
 \item The   measurements   of  the    magnetometers  closer   to  the
       interpolation  points have larger  weights. For  instance, when
       the field is  estimated at the position of TM1,  to which M4 is
       the closest magnetometer, the  value it measures is the largest
       contributor to the interpolated field in TM1. At the same time,
       M1 and M3 being nearly equidistant from both test masses, their
       weights are almost identical.
\end{itemize}

\subsection{Gradient interpolation}
\label{chap.5.2}

The magnetic  field gradient can also be  estimated.  The 9~components
$\partial B_i/\partial B_j$ of the gradient are not independent, since
they must  verify Eqs.~(\ref{eq.4}),  which reduce their  number to~5.
The    remaining~4    components     can    be    easily    calculated
thereafter.  Another option is  to estimate  the~9~gradient components
regardless of the previously  mentioned constraint, in which case they
are  actually found  not to  satisfy them.  Discrepancies  are however
within the estimation error range, so we do not adopt this option here
as it is slightly more cumbersome due to the correspondingly increased
complexity of the network.

\begin{figure}[t]
\centering
\includegraphics[width=0.8\columnwidth]{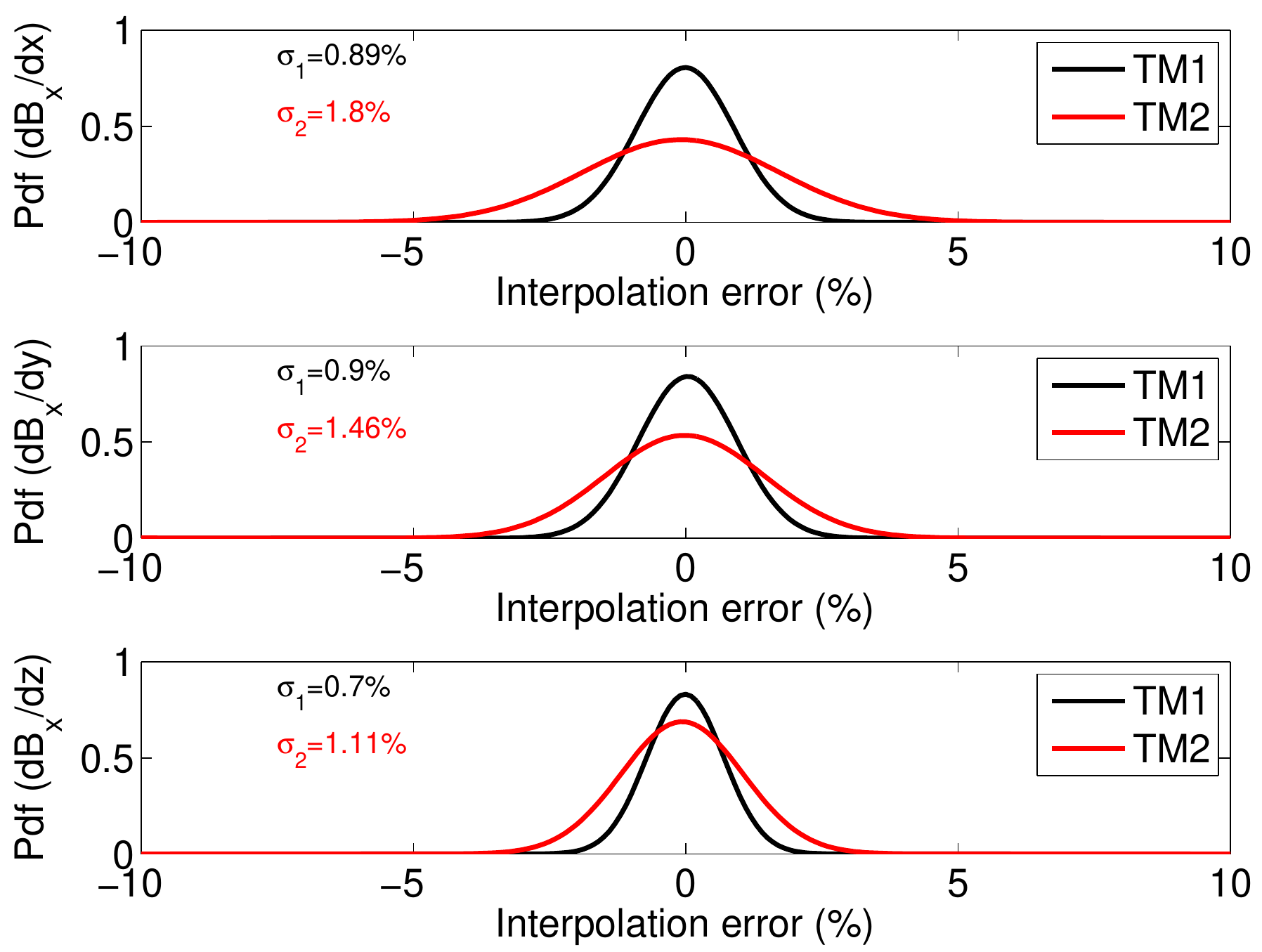}
\caption{Probability density  function of the  errors distribution for
         the  three components of  $\Nabla B_x$.  From top  to bottom:
         $\partial  B_x/\partial  x$,  $\partial B_x/\partial  y$  and
         $\partial  B_x/\partial  z$  at  the positions  of  the  test
         masses.  Errors are  given  in percentages,  the black  lines
         corresponding to TM1, and the red ones to TM2.
\label{fig.6}}
\end{figure}

Results  on gradient  estimation are  shown in  Figure~\ref{fig.6} for
$\Nabla B_x$  at the positions of  both TMs. As can  be observed, they
are  also  within  much  more  acceptable  margins  than  the  earlier
interpolation approach could possibly produce.  It is to be noted that
no  apparent  or  easily  deductible physical  relationship  is  found
between  the estimated  gradient at  the test  mass positions  and the
magnetometer inputs, in contrast with what we have found for the field
estimation.

\subsection{Statistical analysis}
\label{chap.5.3}

\begin{table}
\caption{\label{tab:table2}Statistical properties of the distribution
         of errors of the interpolated magnetic field.}
\begin{ruledtabular}
\begin{tabular}{lrrr}
& $\sigma$ & $\gamma_1$ & $\gamma_2$ \\
\hline
& \multicolumn{3}{c}{Multipole interpolation} \\
\cline{2-4}
$B_x$(TM1)      & 130.7583 & -0.2782 & 19.3869\\
$B_x$(TM2)      & 128.3601 & -0.1009 & 21.4974\\
$|\bf{B}|$(TM1) & 105.5386 & -3.6770 & 29.7343\\
$|\bf{B}|$(TM2) & 102.1037 & -4.4770 & 38.0686\\
\hline
& \multicolumn{3}{c}{Neural network interpolation} \\
\cline{2-4}
$B_x$(TM1)      &   1.5204 & -0.0028 &  2.7746\\
$B_x$(TM2)      &   1.6260 & -0.0008 &  2.8626\\
$|\bf{B}|$(TM1) &   1.4464 & -0.1014 &  2.9440\\
$|\bf{B}|$(TM2) &   1.3682 & -0.0969 &  2.9905\\
\end{tabular}
\end{ruledtabular}
\end{table}

In Table  \ref{tab:table2} we present a statistical  comparison of the
properties of  the distribution of interpolated  magnetic fields.  For
the sake of conciseness we only list the statistical properties of the
interpolated  modulus and  $x$-component  of the  magnetic field.   In
particular,  we   show  the  standard  deviation   ($\sigma$)  of  the
interpolating  errors for  both  the multipole  interpolation and  the
neural network estimate, the skewness of the distribution ($\gamma_1$)
and the  corresponding kurtosis ($\gamma_2$). Clearly,  and as already
mentioned, the  interpolating errors  are very large  for the  case in
which a  multipole interpolating scheme  is used, as clearly  shown by
the  very large standard  deviation obtained  when using  this method.
Also interesting to note is that  for the case of the $x$-component of
the magnetic  field both methods yield distributions  which are almost
symmetrical.   However this is  not the  case for  the modulus  of the
magnetic  field  when  the  multipole interpolating  method  is  used.
Finally, the  kurtosis of the  multipole interpolation is  very large,
revealing a  large number  of outliers.  All in all,  a look  at Table
\ref{tab:table2} reveals that the  neural network method presents much
better statistical properties than the multipole interpolation.


\section{Conclusions}
\label{chap.6}

The magnetic diagnostics  sensor set in the LTP is  such that to infer
the magnetic field  and gradient at the positions of  the TMs based on
the  readouts of  the  magnetometers  is far  from  simple.  The  more
standard interpolation  scheme, based on a multipole  expansion of the
magnetic  field  inside  the   {\sl  LCA}  volume,  cannot  go  beyond
quadrupole  order  which,  in  practice,  means  that  just  a  linear
approximation can be done, due  to the reduced number of magnetometers
available. This grossly fails to produce reliable results, with errors
exceedingly  large.    This  has  motivated  our   search  for  better
alternatives. Artificial neural networks have been presented as a more
elaborate, non-linear procedure to  estimate the required field values
at the  TM positions.  In this  paper we have  presented results which
very significantly improve the  performance of the multipole expansion
technique by almost  two orders of magnitude. This  a very encouraging
outcome which points to the use of the neural networks as the baseline
tool to analyse {\sl LTP} magnetic data.

One of  the main problems  of using the  neural network to  assess the
magnetic  field at  the positions  of  the test  masses is  to find  a
training process  adequate to the  set of data that  the magnetometers
will deliver  in flight. This  underlines the need to  characterize on
ground to our best ability  the magnetic field distribution across the
{\sl LCA} for as many as possible foreseeable working conditions, both
regarding DC  and fluctuating values. Reliable information  on this is
essential for  a meaningful assessment  of magnetic noise in  the {\sl
LTP}.  However, the neural  network analyses  presented in  this paper
only apply  to static fields. What  they actually show  is that neural
networks work  very well  ($\sim$\,10\,\% accuracies) no  matter which
the source dipole configuration is. A different issue, which is beyond
the scope  of this paper and  it is currently  under investigation, is
how to  deal with  time series of  magnetometer readouts, which  is of
course  the  kind of  data  the  satellite  will transmit  to  ground.
Features such as trends,  field fluctuations,\ldots will likely happen
during mission  operations, and the  neural network algorithm  must be
trained to properly deal  with them. Preliminary results indicate that
the  network  is able  to  deal with  moderate  trends  and levels  of
fluctuations, but  further effort  is needed to  explore alternatives,
e.g.\ self-adaptability,  which will make more  robust the performance
of the system.

\begin{acknowledgments}
Support   for   this  work   came   from   grants  ESP2004-01647   and
AYA08--04211--C02--01  of  the   Spanish  Ministry  of  Education  and
Science.  Part of this work was also supported by the AGAUR.
\end{acknowledgments}


\begin{thebibliography}{99}

\bibitem{bib:LISA}
The \textsl{LISA} International Team 2008 ESA-NASA Report LISA-SCRD-Iss5-Rev1
(2008).

\bibitem{bib:trento} 
S Vitale {\sl Science Requirements and Top-level Architecture Definition
for the Lisa Technology Package (\textsl{LTP}) on Board LISA Pathfinder
(SMART-2)}, Report no. LTPA-UTN-ScRD-Iss003-Rev1 (2005).

\bibitem{bib:ere2006} Ara\'ujo H \emph{et al.}  J. of Phys.
Conf. Ser. {\bf 66}, 012003 (2007).

\bibitem{bib:wealthy} Private communication from David Wealthy (Astrium
Stevenage, UK).

\bibitem{bib:ntcs} J. Sanju\'an \emph{et al.} Rev. of Sci. Inst.
{\bf 79} 084503 (2008).

\bibitem{bib:DDS_LTP}
P Canizares \emph{et al.} Class. \& Quantum Grav. \textbf{26}, 094005 (2009).

\bibitem{bib:jackson}
J D Jackson, {\sl Classical Electrodynamics} John Wiley \& Sons, 3rd edition 
(1999)

\bibitem{bib:Kecman}
V Kecman {\sl Learning and soft computing} The MIT Press, 1st edition (2001).

\bibitem{bib:neuralPaper}
C Serpico and C Visone IEEE Trans. on Magnetics, \textbf{34} 623-628
(1998).

\bibitem{bib:neuralPrunning}
R Reed {\sl Prunning algorithms --- A survey} IEEE Trans. on Neural
Networks, \textbf{4}  740-747 (1993).

\end{thebibliography}
\end{document}